\documentclass[aps,twocolumn, showpacs, floatfix, bibnotes]{revtex4}

\usepackage{graphicx}%
\usepackage{dcolumn}
\usepackage{color}
\usepackage{amsmath}
\usepackage{here}

\begin{document}
\begin{sloppy}
\newcommand{\be}{\begin{equation}}
\newcommand{\ee}{\end{equation}}
\newcommand{\bea}{\begin{eqnarray}}
\newcommand{\eea}{\end{eqnarray}}
\newcommand\bibtranslation[1]{English translation: {#1}}
\newcommand\bibfollowup[1]{{#1}}

\newcommand\pictc[5]{\begin{figure}
                       \centerline{
                       \includegraphics[width=#1\columnwidth]{#3}}
                   \protect\caption{\protect\label{fig:#4} #5}
                    \end{figure}            }
\newcommand\pict[4][.9]{\pictc{#1}{!tb}{#2}{#3}{#4}}
\newcommand\rpict[1]{\ref{fig:#1}}

\newcommand\leqt[1]{\protect\label{eq:#1}}
\newcommand\reqtn[1]{\ref{eq:#1}}\newcommand\pictFig[1]{\pagebreak \centerline{
                   \includegraphics[width=\columnwidth]{#1}}
                   \vspace*{2cm}
                   \centerline{Fig. \protect\addtocounter{Fig}{1}\theFig.}}
\newcommand\reqt[1]{(\reqtn{#1})}

\newcounter{Fig}

\title{Complete bandgaps in one-dimensional left-handed periodic structures}

\author{Ilya V. Shadrivov}
\author{Andrey A. Sukhorukov}
\author{Yuri S. Kivshar}

\affiliation{Nonlinear Physics Centre, Research School of
Physical Sciences and Engineering, Australian National University,
Canberra ACT 0200, Australia}

\begin{abstract}
Artificially fabricated structures with periodically modulated
parameters such as photonic crystals offer novel ways of
controlling the flow of light due to the existence of a range of
forbidden frequencies associated with a photonic bandgap. It is
believed that modulation of the refractive index in all three
spatial dimensions is required to open a complete bandgap and
prevent the propagation of electromagnetic waves in all
directions. Here we reveal that, in a sharp contrast to what was
known before and contrary to the accepted physical intuition, a
one-dimensional periodic structure containing the layers of
transparent left-handed (or negative-index) metamaterial can trap
light in three-dimensional space due to the existence of a
complete bandgap.
\end{abstract}


\maketitle

Photonic crystals are artificial materials with a periodic
modulation in the dielectric constant which can create a range of
forbidden frequencies called a photonic
bandgap~\cite{Joannopoulos:2997-143:NAT}. Photons with frequencies
within the bandgap cannot propagate through the medium. This
unique feature can alter dramatically the properties of light,
enabling control of spontaneous emission in quantum devices and
light manipulation for photonic information
technology~\cite{Lodahl:2004-654:NAT}. Photonic bandgap structures
can also be found in nature, and they explain the color diversity
of some of the living creatures~\cite{Vukusic:2003-852:NAT}.

Complete two-dimensional (2D) and three-dimensional (3D) bandgaps
can be realized in photonic crystals, where the refractive index
is periodically modulated in two or three dimensions,
respectively~\cite{Joannopoulos:2997-143:NAT}. Such modulation is
necessary to satisfy the Bragg condition simultaneously for all
propagation directions, requiring that phase accumulation per
period is close to a multiple of $\pi$, so that the waves
reflected at different interfaces between the regions with low and
high refractive indexes interfere constructively and wave
propagation is prohibited for any incidence angle. Manufacturing
of 3D photonic crystals still remains a technological challenge
due to the requirements of large index contrast and high
fabrication precision.

The simplest periodic structure, both in geometry and
manufacturing, is a one-dimensional stack of two types of layers
which differ in the dielectric
constant~\cite{Yeh:1988:OpticalWaves}. However, such structures
{\em may only possess partial bandgaps} for certain ranges of
propagation directions. Here we study the scattering properties of
one-dimensional periodic structures containing layers made of the
so-called {\em left-handed metamaterials} (LHMs)---artificially
created composites which are characterized by simultaneously
negative dielectric permittivity and magnetic permeability. Such
materials are transparent and can bend light in the opposite
direction to normal reversing the way in which refraction usually
works~\cite{Smith:2004-788:SCI}. We demonstrate that specially
designed {\em one-dimensional structures with negative refraction
may exhibit a complete three-dimensional bandgap}. First, we show
that a layered structure made of two alternating layers of
left-handed materials and conventional dielectrics can exhibit a
complete two-dimensional bandgap, i.e. it does not support any
propagating TE or TM waves and is therefore opaque for {\em any
angle of propagation} in the plane. As a direct consequence of such
bandgap, the radiation of waves with given polarization by any
source placed {\em anywhere inside this structure} is prohibited, and the two-dimensional density of states (DOS)~\cite{Busch:1998-3896:PRE} is zero. This result is in sharp contrast to directional reflection~\cite{Winn:1998-1573:OL,
Chigrin:1999-25:APA, Hart:2002-510:SCI} in conventional layered
dielectric structures, which reflect only electromagnetic waves launched from air or a low-index medium. In the periodic structures with usual dielectrics a source dipole can emit radiation of both polarizations, indicating that the
complete gap is absent and propagating TE and TM modes are always
present. The out-coupling of radiation from a periodic structure can vanish only at certain interfaces (see Ref.~\cite{Zurita:2002-39901:PRE+} and references therein), whereas the radiation along the layers, as well as DOS, is never zero. We reveal the physical effects which lead to such fundamental differences between periodic structures with conventional and left-handed layers. In the final part of the paper, we suggest a design of a {\em one-dimensional periodic structure consisting of
three alternating layers} made of conventional dielectric and
left-handed materials which possesses {\em a complete
three-dimensional bandgap}.

We consider a one-dimensional periodic structure created by layers
(with the thickness $d_1$ and $d_2$) of two different materials
with dielectric permittivities $\epsilon_{1,2}$ and magnetic
permeabilities $\mu_{1,2}$, respectively, as shown in
Fig.~\rpict{fig1}. First, we study the propagation of the
TE-polarized electromagnetic waves which have the component of the
electric field parallel to the layers ($E = E_y$); all results can
be easily generalized for the case of the TM polarized waves. We
consider the wave propagation in the ($x,z$) plane characterized
by the wavevector ${\bf k}=(k_x,0,k_z)$. The TE-polarized waves
are described by the linear Helmholtz-type equation for the
electric field component,
\be \leqt{Helm}
   \Delta E
   + \frac{\omega^2}{c^2} n^2(x) E
   - \frac{1}{\mu(x)} \frac{\partial \mu}{\partial x}
   \frac{\partial E}{\partial x}
   = 0,
\ee
where $\Delta = \partial^2 / \partial x^2 + \partial^2 / \partial
z^2$ is the two-dimensional Laplacian, $n^2(x) =
\epsilon(x)\mu(x)$ is the square of refractive index. In a
one-dimensional periodic structure the propagating waves have the
form of {\em Bloch modes}, for which the electric field amplitudes
satisfy the periodicity condition, $E(x+\Lambda, z) = E(x, 0)
\exp(i K_b + i k_z z)$, where $\Lambda = d_1 + d_2$ is the period
of the structure. Here $K_b$ is the dimensionless Bloch wave
number which defines the wave transmission across the layers, and
its dependence on the wavevector component along the layers
($k_z$) can be found explicitly for two-layered periodic
structures (see, e.g., Refs.~\cite{Yeh:1988:OpticalWaves,
Li:2003-83901:PRL}),
\bea \leqt{TransferM}
  \begin{array}{l} {\displaystyle
    2 \cos(K_b) = {\rm Tr}(M) =
     2 \cos( k_{1x} d_1 ) \cos( k_{2x} d_2 )
   } \\*[9pt] {\displaystyle \qquad
     - \left(   \frac{k_{2x} \mu_1}{k_{1x} \mu_2}
              + \frac{k_{1x} \mu_2}{k_{2x} \mu_1} \right)
     \sin( k_{1x} d_1 ) \sin( k_{2x} d_2 ) .
   } \end{array}
\eea
Here ${\rm Tr}(M)$ is the trace of the transfer matrix $M$
characterizing the wave scattering in a periodic
structure~\cite{Yeh:1988:OpticalWaves}, $k_{jx} = k_j (1 - k_{z}^2
/ k_j^2)^{1/2}$ are the $x$-components of the wavevector in the
first ($j=1$) and second ($j=2$) media, and $k_j = \omega n_j / c$
are wavenumbers in each media with refractive indexes $n_j$. For
completeness, we mention that the dispersion relation of the TM
polarized waves is obtained by replacing $\epsilon \Leftrightarrow
\mu$ in Eq.~\reqt{TransferM}.

Solutions of the dispersion relation~\reqt{TransferM} with both
real $k_z$ and $K_b$ correspond to Bloch waves which can propagate
through the periodic structure, whereas complex $k_z$ or $K_b$
indicate the presence of {\em bandgaps} in the spectrum where the
wave propagation is prohibited. A complete bandgap occurs if for
all real $k_z$, the $K_b$ remains complex. It was recently
shown~\cite{Nefedov:2002-36611:PRE, Li:2003-83901:PRL} that novel
partial bandgaps can appear in structures made of alternating
layers of LHM and normal dielectrics when the condition of zero
average refractive index is satisfied for particular propagation
angles ($k_z$), $k_{1x} d_1 + k_{2x} d_2 = 0$, which is possible
because $k_x$ is positive in conventional dielectrics and it is
negative in left-handed materials. However, we find that this
requirement is neither sufficient nor necessary to obtain complete
bandgaps.

\pict[0.8]{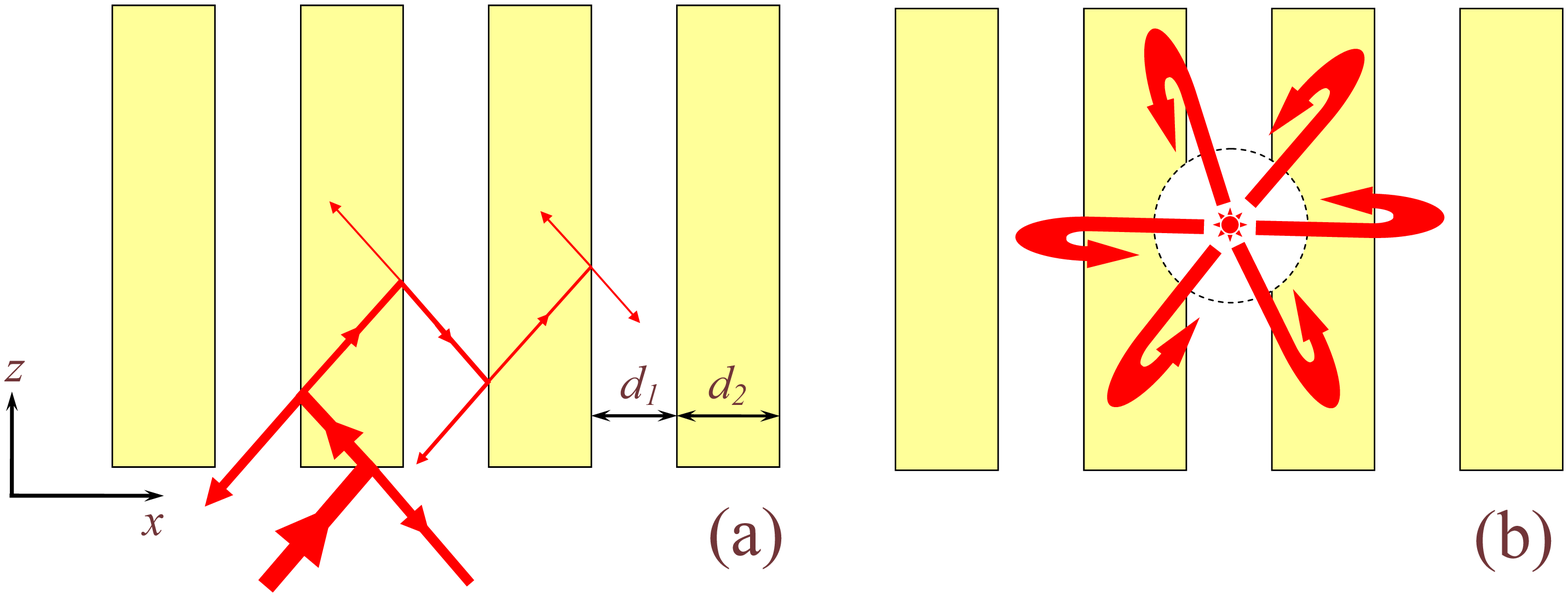}{fig1}{Schematic of the left-handed periodic
structure with (a) the ray diagram for the end-fired wave
propagation and (b) suppression of radiation of a local source
placed inside.}

In order to emphasize the importance of our findings presented
below, first we recall the basic physics which explains why
one-dimensional periodic structures containing materials of the
same type (i.e., normal dielectrics) do not possess a complete
three-dimensional bandgap. Analyzing the effects associated with
the wave scattering in Bragg
gratings~\cite{Russel:1995-585:ConfinedElectrons}, we come to the
conclusion that the only phenomenon which always allows for the
wave propagation in the 1D dielectric periodic structures, and
which cannot be suppressed by a choice of the structural
parameters, is the {\em wave guiding by optically dense layers}.
Indeed, it is well known that a dielectric waveguide with the core
made of an optically dense medium always supports a fundamental
mode. However, as was shown
recently~\cite{Shadrivov:2003-57602:PRE}, the fundamental mode can
be absent if the core is made of LH metamaterial. And it is this
property of a LH waveguide that allows us to introduce a novel
type of one-dimensional periodic structures with a complete
bandgap.

{\em Why does the usual dielectric waveguide always support the
fundamental mode, and why a metamaterial waveguide does not?} The
condition for the guided waves to exist, defined by the dispersion
relation for the modes in a slab waveguide, has a simple physical
meaning: the round-trip accumulation of phase due to wave
propagation across the layer, $2\phi_{\rm prop}$, including the
phase retardation upon the total internal reflection, $2\phi_{\rm
refl}$, should be equal to a multiple of $2\pi$. The phase change
due to the total internal reflection is {\em negative} for both
types of waveguides, and depending on the angle of incidence it
varies from $0$ to $-\pi$. A difference between the conventional
and LHM waveguides appears due to the phase accumulated by the
wave propagating across the layer. In usual dielectrics, the wave
is {\em forward}, i.e. the phase accumulated along the direction
of energy flow is positive. As a result, there always exists an
angle of wave propagation, such that the total phase change
vanishes, and at least one mode always exists in a dielectric
waveguide. In a LHM waveguide, the wave is {\em backward}, and the
phase change $\phi_{\rm prop}$ is negative. Then, one can choose
the parameters in such a way that for all angles of wave
propagation (greater than the angle of the total internal
reflection), the total phase change is $-4 \pi < \phi =
2(\phi_{\rm refl}+\phi_{\rm prop}) < -2 \pi$, and no guided modes
exist.

\pict{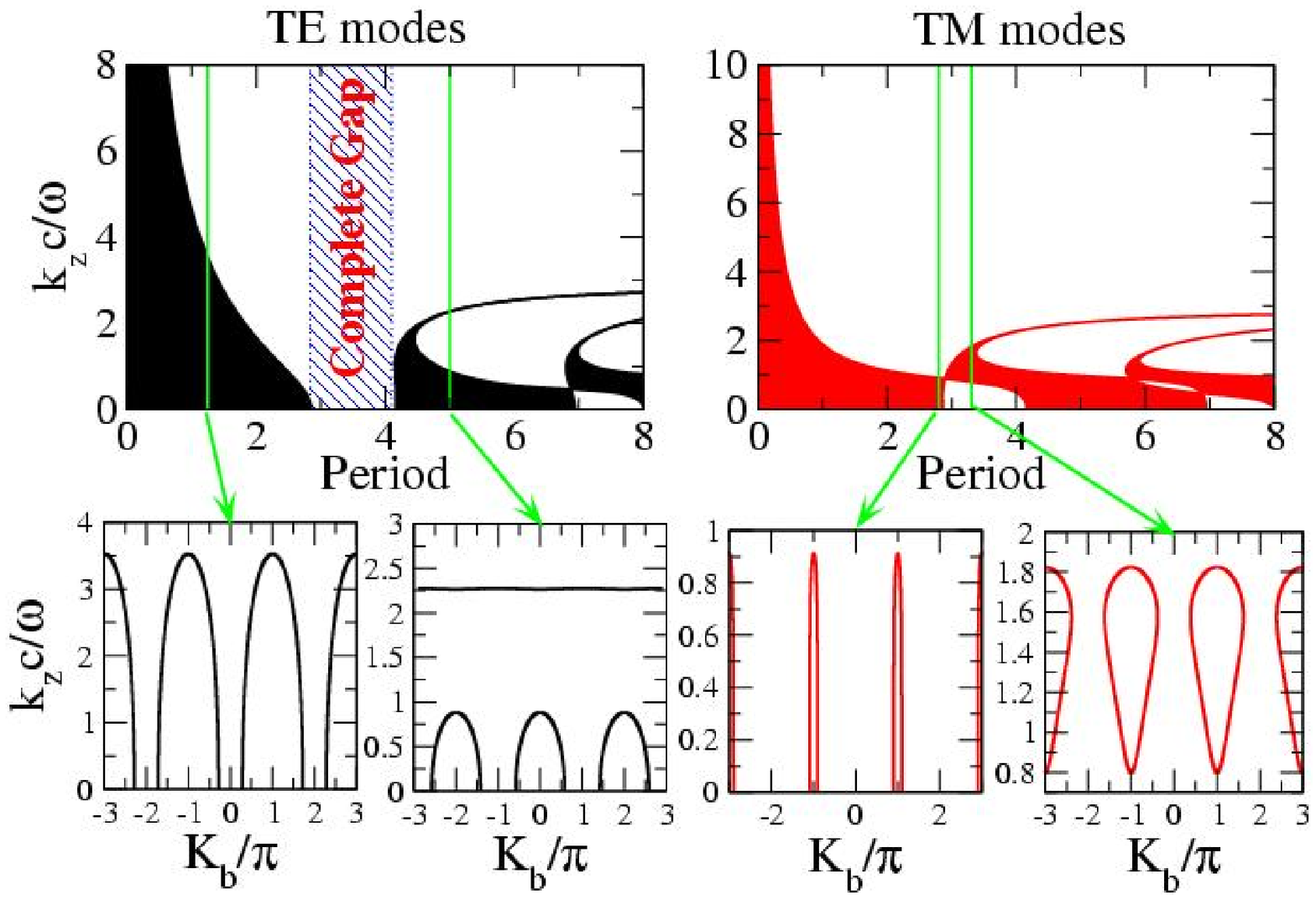}{KyVsSize}{ (top) Shaded regions mark ranges of
longitudinal wavenumbers vs. the structure period [normalized to
$\lambda/(2\pi)$] corresponding to TE (left) and TM (right)
polarized electromagnetic waves that can propagate through the
structure. (bottom) Dispersion diagrams of the Bloch waves for
particular polarizations and periods, as indicated by arrows.}

Based on the results presented above we can construct the
one-dimensional periodic structure which possesses a complete
bandgap for one polarization. Indeed, we need to choose the layer
thickness such that the {\em guided modes are absent} in the
waveguides formed by the layers in the structure, and by varying
the material parameters we should {\em avoid the transmission
resonances}~\cite{Russel:1995-585:ConfinedElectrons}. Our analysis
shows that it is indeed possible to find such structures and, for
example, the complete gap appears for the following set of
parameters: $\epsilon_1=\mu_1 = 1$, $\epsilon_2 = -6$, $\mu_2 =
-1.38$, $d_1^{(0)} = 1.5\lambda/2\pi$, $d_2^{(0)} =
1.4\lambda/2\pi$, where $\lambda = 2 \pi c / \omega$ is the vacuum
wavelength. Then we study the dependence of the bandgap spectrum
on the structure period for a fixed ratio $d_1/d_2 =
d_1^{(0)}/d_2^{(0)}$, and Fig.~\rpict{KyVsSize}(top) shows the
bands (colored) where the wave propagation is possible. For the TE
modes [see Fig.~\rpict{KyVsSize}(left, top)], there exists a range
of periods for which the propagation is completely prohibited for
all possible $k_z$, since the corresponding $K_b$ are complex.
This means, in particular, that if we consider an end-fire
generation problem and launch the waves along the layers towards
the structure, {\em they will be completely reflected}, as
schematically shown in Fig.~\rpict{fig1}. {\em Such a regime is
impossible for any type of the conventional dielectric gratings.}

However, there is no complete bandgap for the TM-polarized waves
propagating in the same structure. In the optimal case with
$d_{1,2} \approx d_{1,2}^{(0)}$, we have only one angle of
propagation possible (i.e. the only value of $k_z$ with real
$K_b$). This is {\em the Brewster angle} for which there is no
reflection of TM waves at the interfaces. From the electromagnetic
duality principle we find that taking the structure with LHM
metamaterial characterized by $\epsilon_2^{\rm new}=\mu_2$ and
$\mu_2^{\rm new} = \epsilon_2$, we can obtain a complete bandgap
for the TM polarized waves. Most remarkably, the complete bandgap
for each polarization exists for rather broad ranges of the
structure parameters, see Fig.~\rpict{param_plane}. The regions
for the complete TE and TM bandgaps are symmetric with respect to
the line $\epsilon_2 = \mu_2$, and this is a consequence of the
electromagnetic duality.

\pict[0.6]{fig03}{param_plane}{ Regions of  parameters for
which the complete TE (black) and complete TM (red) bandgaps
exist. The thickness $d_1^{(0)}$ and $d_2^{(0)}$ of the LH and RH
layers are defined in the text.}

\pict{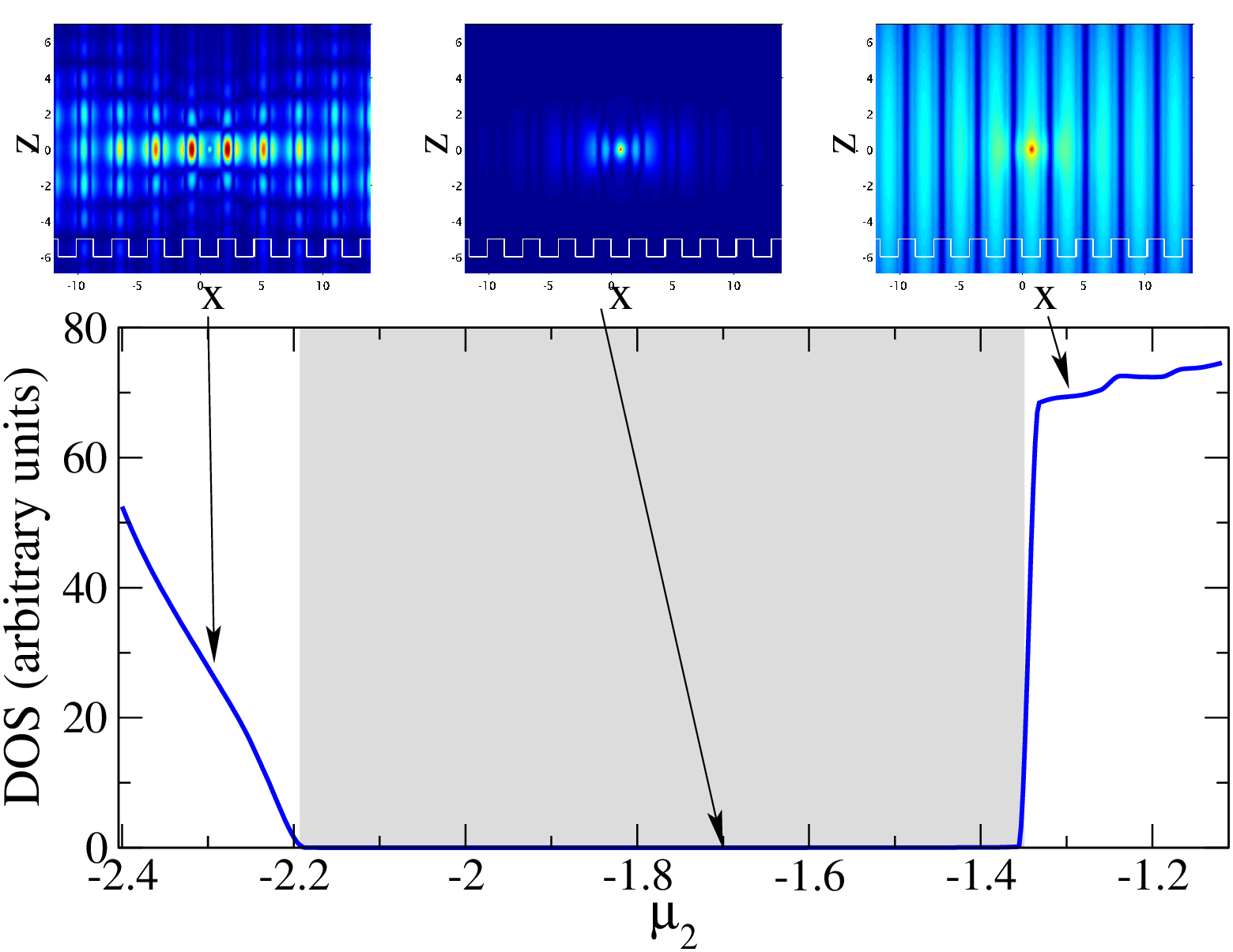}{w_mu}{ Dependence of the two-dimensional density of states on the magnetic permeability of LHM. The top insets show the structure of the Green's function inside the bandgap (middle), and outside the bandgap (left and right). The source position is $x_0 = z_0 = 0$.}
One-dimensional structures with a complete bandgap for one of the
polarizations can be used to form an electromagnetic cavity. To
study the main features of the wave localization due to the presence of a complete band-gap, we analyze the field of a line current $J$ running along the $y$-axis inside the structure at the position ${\bf r}_0 = (x_0,z_0)$ in the ${\bf r} = (x,z)$ plane. It follows from the Maxwell's equations that the electric field can be expressed as $E(x,z)=i \omega J \mu(x_0, z_0) G(x,z) /c^2$, where $G$ is the Green's function found as a solution of the following equation,
\be \Delta G + \frac{\omega^2}{c^2}n^2(x) G
   - \frac{1}{\mu(x)} \frac{\partial \mu}{\partial x}
   \frac{\partial G}{\partial x}
  = 4\pi\delta({\bf r} - {\bf r}_0).
\ee
The total emitted power per unit length of the line current $J$ is
\be
W = -\frac{\omega J^2 \mu(x_0,z_0)}{2 c^2} {\rm Im} 
            \left[{G(x_0,z_0)} \right],
\ee
and this quantity is proportional to the local density of states (LDOS)~\cite{Sheng:1995:Scattering}. The density of states (DOS) of the structure, which is an integral of LDOS over the Brillouin zone, characterizes the radiation efficiency of multiple sources located at different positions. The DOS becomes zero only if radiation in any direction in the plane is prohibited, indicated the presence of a complete 2D band-gap. We plot the dependence of the DOS on the magnetic permeability of negative index material $\mu_2$ for fixed $\epsilon_2 = -6$ and $d_{1,2} = d_{1,2}^{(0)}$ in Fig.~\rpict{w_mu}, which clearly demonstrates that a two-dimensional band-gap exists for the TE-polarized waves within a certain range of media parameters. Within the bandgap region the radiation is suppressed and the Green's function is exponentially localized for any source position [see Fig.~\rpict{w_mu}(top, middle)]. Outside the bandgap, different propagating Bloch modes are excited, and the Green's function is not localized [see Fig.~\rpict{w_mu}(top, left and right)].

\pict{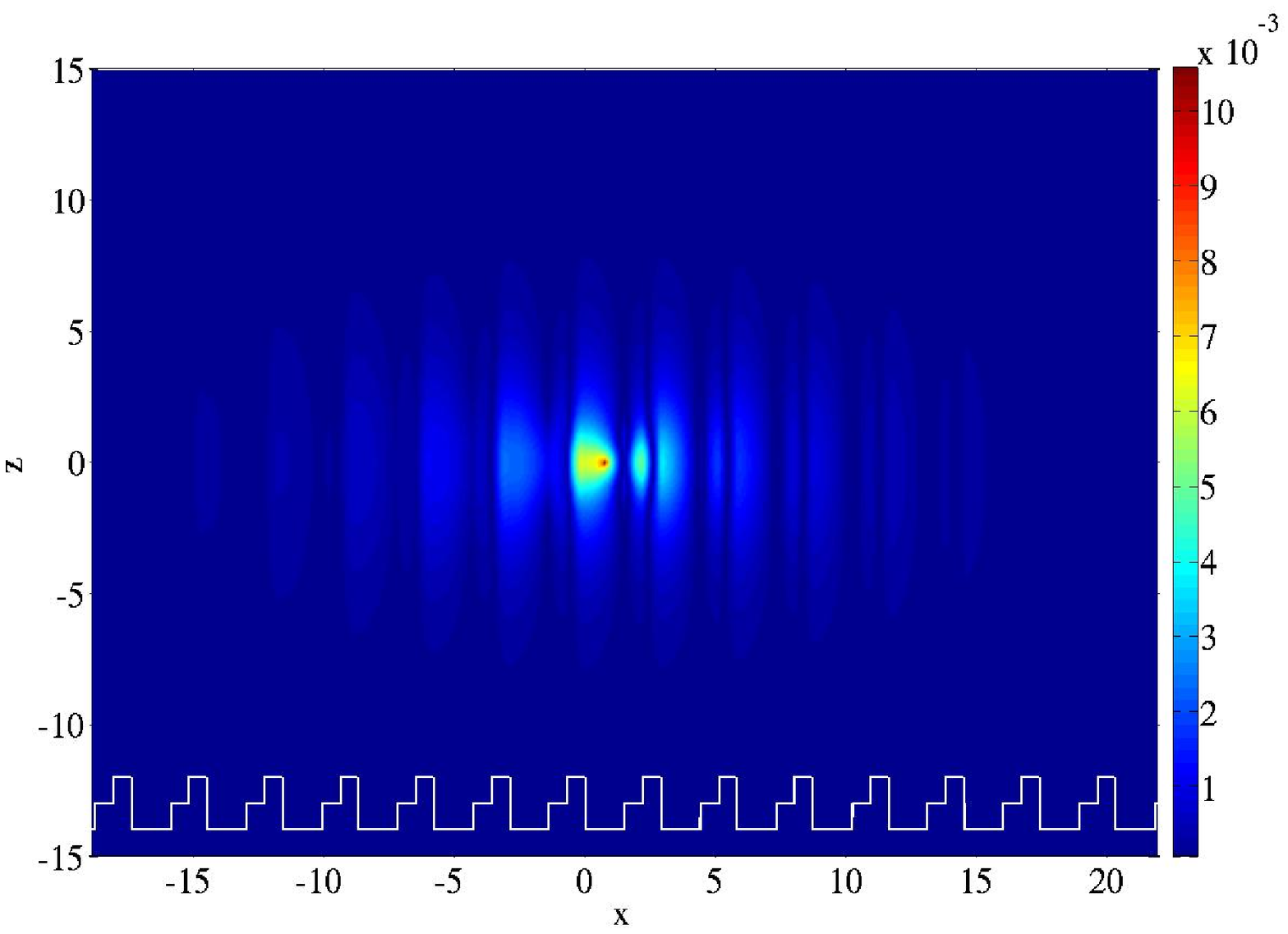}{3layerGreen}{The Green's function of a
one-dimensional three-layer periodic structure possessing an
absolute bandgap.}

After the comprehensive study of the two-layer periodic systems
and the properties of the complete bandgaps supported by
one-dimensional hybrid structures, we are able to suggest the case
when the complete bandgap may appear for both polarizations thus
allowing the existence of the absolute bandgap. Indeed, to do this
we should consider more sophisticated case of a three-layer
periodic structure in order to suppress the conditions for the
existence of the Brewster angle which prevents us from creating a
complete bandgap in two-layer structures. The Brewster-angle
transmission resonance can be easily eliminated by introducing a
third layer in the structure, thus allowing the existence of a
complete three-dimensional bandgap for all waves propagating
inside a specially designed one-dimensional structure. To
demonstrate this quite unique property, we choose the structure
with the parameters $\epsilon_1 = \mu_1 = 1$, $\epsilon_2 =
\mu_3$, $\mu_2 = \epsilon_3$, and $d_2 = d_3$. The choice of this
symmetry simplifies the analysis of the structure making it
mathematically more elegant. Indeed, in this case the trace of the
transfer matrix is {\em the same for both polarizations}, and it
can be represented in the form
\be
  \begin{array}{l} {\displaystyle
     {\rm Tr}(M) = 2 \cos{\left( k_{1x} d_1 \right) }
       \cos^2{\left( k_{2x} d_2 \right)}
  } \\*[9pt] {\displaystyle \,
     - \left( \frac{\epsilon_2}{\mu_2} + \frac{\mu_2}{\epsilon_2} \right)
    \cos{\left( k_{1x} d_1 \right) } \sin^2{\left( k_{2x} d_2 \right)}
  } \\*[9pt] {\displaystyle \,
     - \frac{1}{2} (\epsilon_2 + \mu_2)
        \left( \frac{k_{1x}}{k_{2x}}
              + \frac{1}{\epsilon_2 \mu_2}\frac{k_{2x}}{k_{1x}} \right)
        \sin{\left( k_{1x} d_1 \right) }\sin{\left( 2 k_{2x} d_2 \right)}.
  } \end{array} \nonumber
\ee
Since the traces of the transfer matrices coincide for both TE and
TM polarizations, the bandgaps will appear in the structure
spectrum for both the polarizations simultaneously. Existence of
the transmission bandgaps for both TE and TM polarized waves in
the same periodic structure indicate the existence of {\em an
absolute bandgap}. As an example, we show that the structure with
the parameters $\epsilon_2 = \mu_3 = -6$, $\epsilon_3 = \mu_2 =
-1.38$, $d_1=1.5 \lambda/ (2 \pi)$, $d_2=d_3=0.7 \lambda/ (2 \pi$)
possesses {\em an absolute three-dimensional bandgap}, and the
Green's function corresponding to this three-layer structure is
presented in Fig.~\rpict{3layerGreen}.

In conclusion, we have revealed a novel and highly nontrivial
property of left-handed metamaterials with negative refraction: A
one-dimensional periodic structure containing layers made of a
left-handed metamaterial can trap light in three dimensions due to
the existence of a complete photonic bandgap. This finding is in a
sharp contrast with the fundamental concepts of the conventional
physics of photonic crystals where complicated structures with
two- and three-dimensional periodicity are required. We believe
that our results suggest new directions for the future
applications of metamaterials for microwaves, Terahertz
frequencies, and visible light as fabrication technologies become
available.

We thank C. Soukoulis, P. Belov, I. Gabitov, S. Mingaleev, V. Shalaev, and H.
Winful for useful discussions.

\end{sloppy}

\end{document}